\documentstyle[12pt,epsfig]{article}
\def\setb@se#1{\baselineskip=#1 \normalbaselineskip=#1}
\setb@se{14pt}
\newcommand{\be}{\begin{equation}}
\newcommand{\ee}{\end{equation}}
\newcommand{\bea}{\begin{eqnarray}}
\newcommand{\eea}{\end{eqnarray}}
\newcommand{\baf}{{\bar f}}
\newcommand{\bet}{{\bar\eta}}
\newcommand{\ov}[2]{{#1\over #2}}
\newcommand{\rh}{r_{\rm h}}
\newcommand{\fh}{{\bar f}_{\rm h}}
\newcommand{\mh}{\mu_{\rm h}}
\newcommand{\ph}{\psi_{\rm h}}

\newcommand{\rc}{r_{\rm c}}
\newcommand{\rp}{r_{+}}
\newcommand{\xh}{x_{\rm h}}

\begin{document}

\begin{titlepage}
\hbox to\hsize{%

  \vbox{%
        \hbox{MPI-PhT/99-60}%
        \hbox{\today}%
       }}

\vspace{3 cm}

\begin{center}
\Large{Gravitational Global Monopoles with Horizons}

\vskip5mm
\large
Dieter Maison

\vspace{3mm}
{\small\sl
Max-Planck-Institut f\"ur Physik\\
--- Werner Heisenberg Institut ---\\
F\"ohringer Ring 6\\
80805 Munich (Fed. Rep. Germany)\\}

\end{center} 
\vspace{20 mm}
\begingroup \addtolength{\leftskip}{1cm} \addtolength{\rightskip}{1cm}

\begin{center}\bf Abstract\end{center}
\vspace{1mm}\noindent
We give arguments for the existence of ``radial excitations'' 
of gravitational global monopoles with any number of zeros of the 
Higgs field and present numerical results
for solutions with up to two zeros.
All these solutions possess a de Sitter like cosmological horizon, outside
of which they become singular.
In addition we study corresponding static ``hairy'' black hole solutions,
representing black holes sitting inside a global monopole core. In
particular, we determine their existence domains as a function of 
their horizon radius $\rh$.
\endgroup
\end{titlepage}
\newpage
Like the well-known 't Hooft-Polyakov monopoles, global monopoles are
topological solitons with a `hedgehog' type Higgs field \cite{Shellard}. 
Whereas the former depend on two mass scales $M_{\rm W}$ and $M_{\rm H}$ --
the masses of the Yang-Mills resp.\ the Higgs field -- the latter may be 
formally obtained by letting $M_{\rm H}$ tend to infinity 
(keeping $M_{\rm W}$ fixed) and performing a suitable rescaling
of the radial coordinate.
In this limit the YM field vanishes and the local gauge invariance of the 
model is lost, only a global SU(2) invariance remaining.
This infinite rescaling also sends the mass of the monopole to infinity.    
When the global monopole is coupled to gravity another mass scale $M_{\rm
Pl}=1/\sqrt G$ enters and one obtains a one-parameter family of static
global monopoles parametrized by the dimensionless ratio 
$\eta=M_{\rm W}/M_{\rm Pl}$ \cite{Barriola,Liebling}.
Although the space-times of these solutions are asymptotically flat they
have a kind of `conical` singularity at spatial infinity 
expressed by a deficit solid
angle $\Delta=4\pi(8\pi\eta^2)$ -- a remnant of their infinite mass in 
flat space.
Nevertheless self-gravitating global monopoles have been 
considered in the literature in
connection with `topological inflation'~\cite{Linde, Vilenkin}.
This is not unreasonable as the driving mechanism of inflation is
an effective cosmological constant provided by the Higgs potential
in the region of `false vacuum' inside the core of the monopole, the
asymptotic
region far from the monopole playing no role for these considerations.
As argued by Linde and Vilenkin static monopoles should undergo
inflation when the parameter $\eta$ exceeds a certain critical value 
$\eta_{\rm c}$
of order one for which the Schwarzschild radius of the monopole 
becomes equal to the actual size of the monopole core as measured by 
$1/M_{\rm W}$. The same type of argument was used before to exclude the 
existence of static monopoles for $\eta>\eta_{\rm c}$,
confirmed by numerical calculations of static solutions \cite{Ortiz,
Weinberg,BFM}. 

\begin{figure}
\hbox to\hsize{
  \epsfig{file=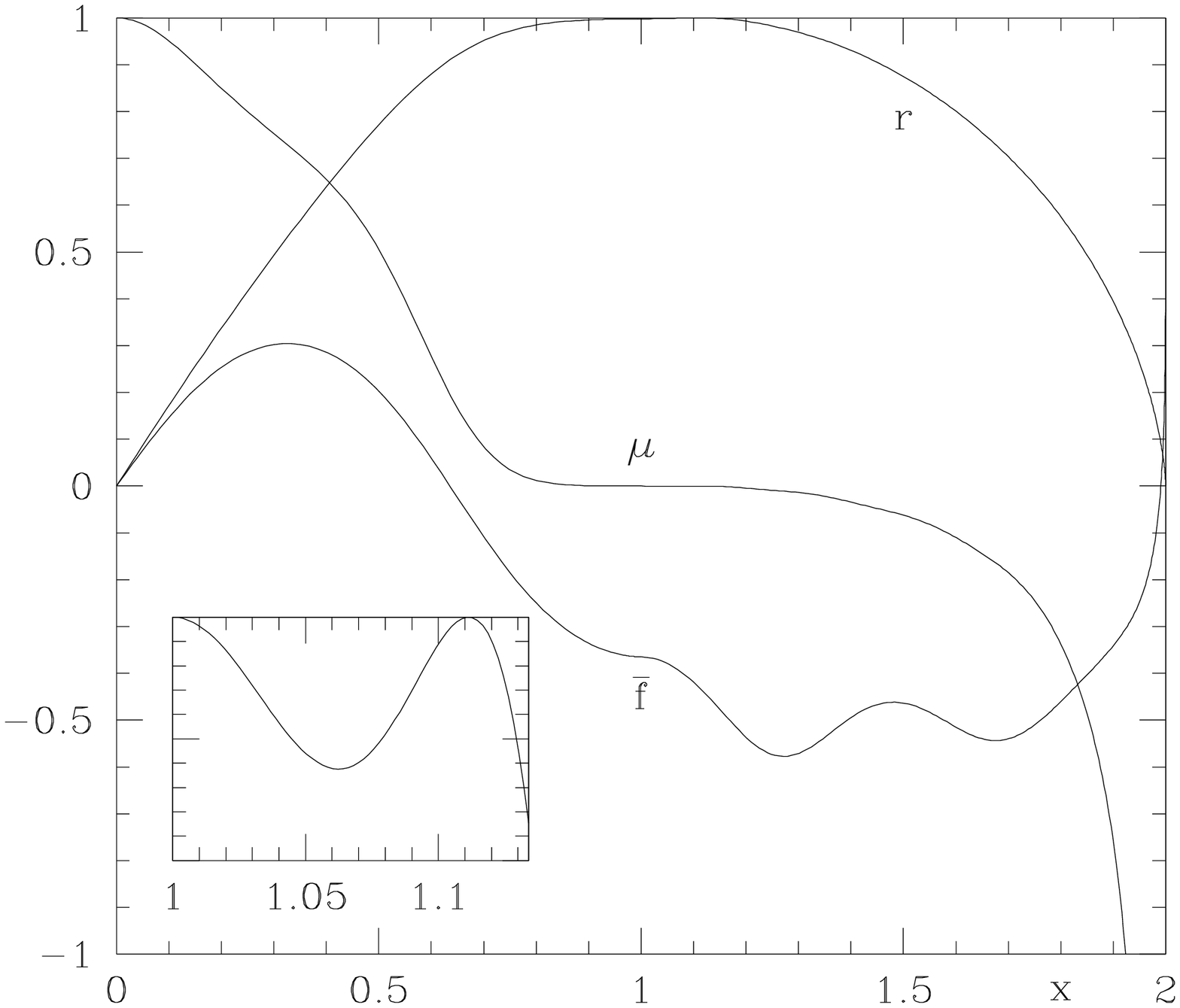,width=0.48\hsize,%
      bbllx=0.5cm,bblly=6.0cm,bburx=20cm,bbury=22.0cm}\hss
  \epsfig{file=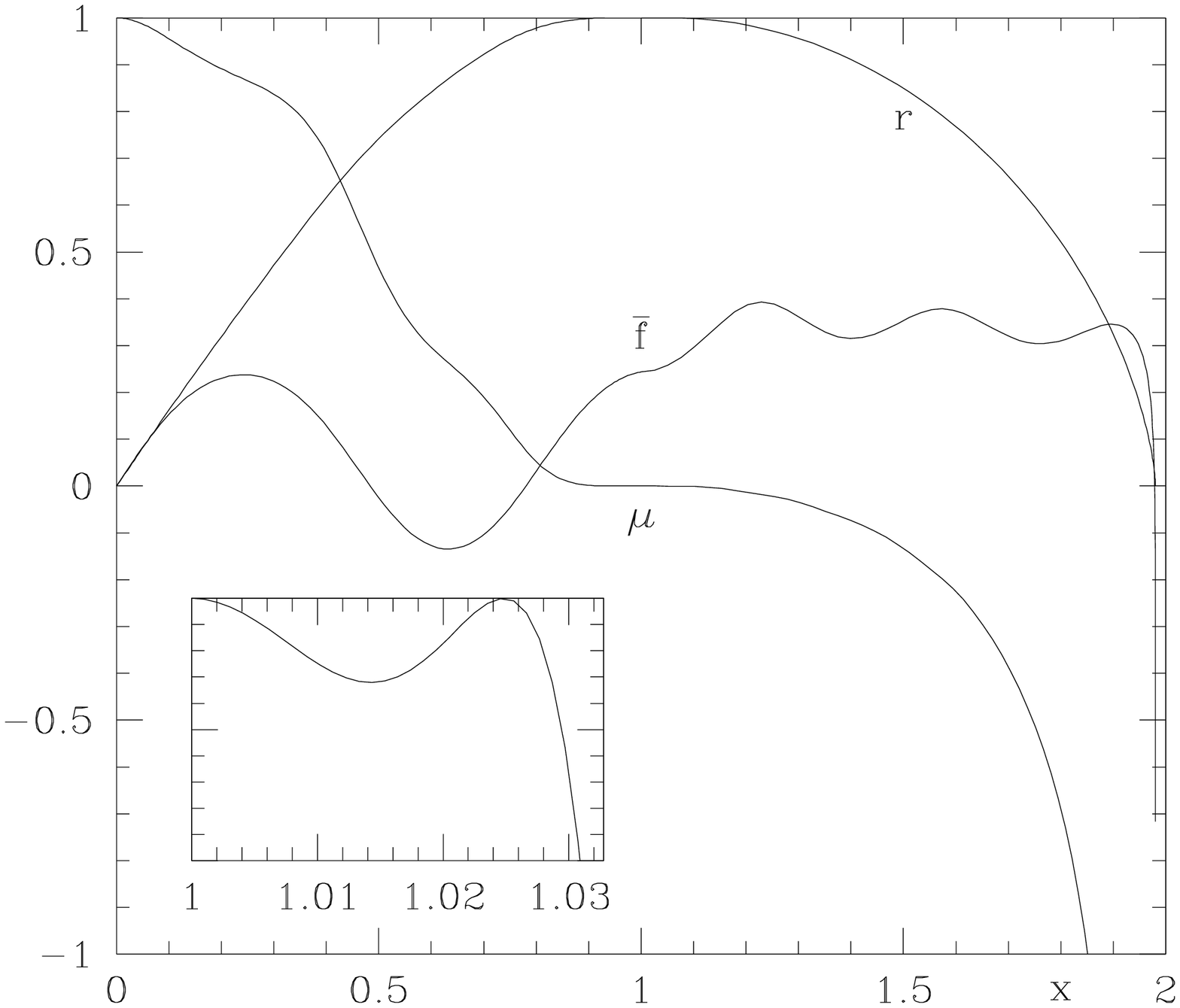,width=0.48\hsize,%
      bbllx=0.5cm,bblly=6.0cm,bburx=20cm,bbury=22.0cm}
  }
\caption[slns]{\label{slns}Global monopoles with 1 and 2 zeros
corresponding to
$\bet^2=0.3$ and $0.13$. The cosmological horizon is situated 
at $x=1$. The inserts show enlargements of $\mu$ by factors 
$5\cdot 10^3$ rsp.\ $10^7$ slightly outside the horizon.
}
\end{figure}

Actually, in the case of global monopoles there is another limit to the
existence of static, asymptotically flat solutions obtained when the deficit
solid angle $\Delta$ becomes equal to $4\pi$, i.e.\ for $\eta=\sqrt{1/8\pi}$. 
Nevertheless, as shown by 
S.~Liebling \cite{Liebling}, there are still static
solutions for larger values of $\eta$ having a de Sitter like cosmological
horizon. Approaching the value $\eta=\sqrt{1/8\pi}$ from above the radius of
the cosmological horizon of these solutions tends to infinity. 
This type of solution exsists up to a maximal value $\eta_{\rm
max}=\sqrt{3/8\pi}$ at which the solution is found to bifurcate with the 
de Sitter solution obtained for vanishing Higgs field. 
As demonstrated in a recent
paper with S.~Liebling \cite{MaisonL}, this bifurcation allows an analytical
derivation of $\eta_{\rm max}=\sqrt{3/8\pi}$ 
related to the existence of a bounded ``zero mode''. 
According to standard arguments \cite{Gelfand} the latter also
implies a change of stability of the de Sitter solution. Whereas the global
monopole is stable on its whole domain of existence $0\leq \eta<\eta_{\rm
max}$ the de Sitter solution is unstable in this domain and stable beyond.
As was shown in \cite{MaisonL} 
there appear new unstable modes of the de Sitter solution, when
$\eta$ decreases at discrete values $\eta_K$ accumulating at $\eta=0$.
It does not require a lot of fantasy to surmise that also these points
could be points of bifurcation with some new kind of global monopole
solutions. Since the corresponding zero modes of the de Sitter solution 
have Higgs fields with $K$ zeros, these new solutions ought to be 
``radial excitations'' of the global monopole with a corresponding number of
zeros. We shall present some numerical evidence that such solutions indeed
exist. Some examples are shown in Fig.(\ref{slns}). 

As in the case of gravitating 't Hooft-Polyakov monopoles 
\cite{Weinberg,BFM} there are also
black hole solutions with a ``global monopole hair''.
For the fundamental solution without zero these have been already
found in \cite{Liebling}. In the present context we not only
describe radial excitations of those, but also determine their
existence region in parameter space, consisting of pairs $(\eta,\rh)$, where
$\rh$ is the radius of the black hole horizon in Schwarzschild coordinates.     

\section{Radially excited gravitating global monopoles}\label{global}

Following the notation of \cite{Liebling}, we put $\phi^a=f(r)\hat r^a$
for the Higgs field and     
\be                                                  \label{metric}
ds^2=-A^2\mu~dt^2+\ov{dr^2}\mu+r^2d\Omega^2
\ee
for the spherically symmetric line element in Schwarzschild coordinates.
The resulting static field equations are (we prefer to use rationalized
units putting $\baf=\sqrt{4\pi}f$, $\bet=\sqrt{4\pi}\eta$, $\lambda=2\pi$ as compared to
$f$ and $\eta$ from \cite{Liebling}) 
\bea\label{fequs}                                                          
\baf'&=&\psi  \label{fequs_f}\\
\psi'&=&\ov{\baf}{r^2\mu}\Bigl[2+\frac{r^2}{2}(\baf^2-\bet^2)\Bigr]-
         \psi\Bigl[\ov{2}{r}+r\psi^2+\ov{\mu'}{\mu}\Bigr]\\ 
\mu'&=&\ov{1-\mu}{r}-r\mu\psi^2-\ov{2\baf^2}{r}-
        \ov{r}{4}\bigl(\baf^2-\bet^2\bigr)^2 \label{fequs_mu}\\
A'&=&r\psi^2 A \label{fequs_a}.
\eea
Solutions with a regular origin obey the boundary conditions
\be                                                  \label{bdcon}
\baf(r)=ar+O(r^3),\quad \psi(r)=a+O(r^2),\quad \mu(r)=1+O(r^2),\quad
A(r)=A_0+O(r^2)
\ee
and are uniquely specified by the choice of $a=\psi(0)$ and $A_0$.

The boundary conditions for black holes at the horizon are
\bea\label{bhbd}                                          
\baf(r)&=&\baf_{\rm h}+O(r-\rh),\quad 
\psi(r)=\psi_{\rm h}+O(r-\rh),\quad\\
\mu(r)&=&\mu'_{\rm h}(r-\rh)+O((r-\rh)^2),\quad
A(r)=A_{\rm h}+O((r-\rh)^2)
\eea
with
\bea\label{bdhor}
\mh'&=&\ov1{\rh}\Bigl[1-2\fh^2-\ov{\rh^2}4\bigl
   (\fh^2-\bet^2\bigr)^2\Bigr]>0\\                  
\ph&=&\ov{\fh}{\rh^2\mh'}\Bigl[2+\ov{\rh^2}2\bigl(\fh^2-
\bet^2\bigr)\Bigr]\,.
\eea
Solutions with any of these boundary conditions stay finite 
for increasing $r$ as long as
$\mu$ is non-zero. However, generically $\mu$ vanishes for some finite value
of $r$.  
Depending on whether $A$ and $\psi$ stay finite or not at the zero
of $\mu$ the geometrical significance of such points is different. 
In the first case 
the solution has a cosmological horizon, in the latter it has an ``equator'',
i.e.\ a maximum of $r$ considered as a metrical function 
(compare eq.(\ref{metric})). The boundary conditions at a cosmological
horizon are the same as at a black hole horizon given above with the only
difference that  $\mu'_{\rm h}<0$.

Solutions with a regular origin resp.\ black hole boundary conditions and a
cosmological horizon can be obtained by fine-tuning the parameter $a$
resp. $\fh$. An important special case is obtained for $a=0$ resp.
$\fh=0$ yielding the de Sitter resp.\ Schwarzschild-de Sitter 
solution with a cosmological constant provided by the Higgs potential
for $\baf\equiv 0$.   
The de Sitter (dS) solution is given by
\be\label{dS}
\mu_{\rm dS}(r)=1-\ov{r^2}{\rc^2} \quad {\rm with}\quad 
     \rc=\ov{2\sqrt{3}}{\bet^2}\,,
\ee
whereas the Schwarzschild-de Sitter (SdS) solution reads
\be\label{SdS}
\mu_{\rm SdS}(r)=1-\ov{r^2}{\rc^2}-\ov{\rh}{r}\Bigl(1-\ov{\rh^2}
                {\rc^2}\Bigr)\, .
\ee

\begin{figure}
\begin{center}
  \epsfig{bbllx=0.5cm,bblly=7.0cm,bburx=20.0cm,bbury=20.0cm,%
          file=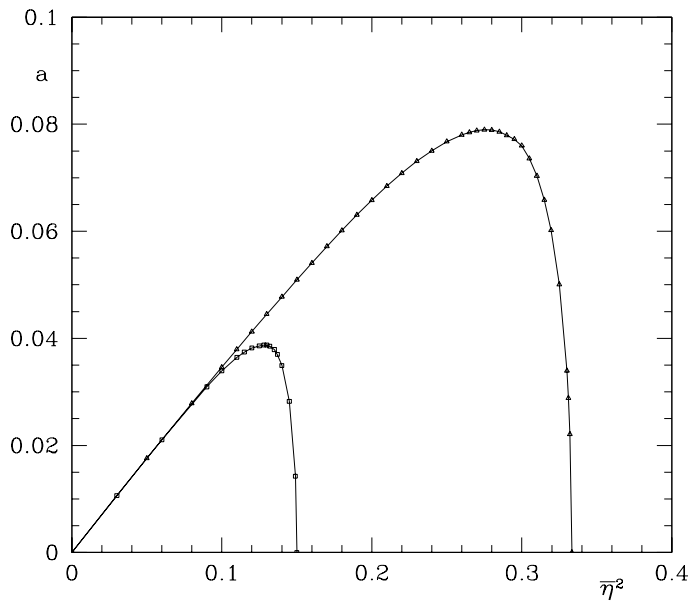,height=6cm}
\end{center}
\caption[acurves]{\label{acurves}$a(\bet^2)$ for the first and second radially
excited global monopoles}
\end{figure}

Obviously the SdS solution goes into the dS solution for $\rh\to 0$.
The cosmological horizon of the SdS solution turns out to be located at
\be\label{chor}
\rp=-\ov{\rh}{2}+\sqrt{\rc^2-\ov{3}{4}\rh^2}
\ee
As was shown in \cite{MaisonL} the de Sitter solution
has bounded ``zero modes'' $\varphi_K$ solving the linearized 
eqs.(\ref{fequs}) in the dS background for the discrete values
\be\label{eta}
\bet_K^2=\ov{3}{(K+2)(2K+1)} \qquad {\rm for}\quad  K=0,1,2,..
\ee                                
At the largest one $\bet_0^2=3/2$ the de Sitter solution bifurcates with 
the global monopole solutions possessing a themselves a cosmological horizon
for $\bet^2>1/2$ \cite{Liebling}. 
The corresponding zero mode is given by
$\varphi_0=\partial\baf/\partial a$, where $a$ is the parameter
characterizing the solution at $r=0$ (compare eq.(\ref{bdcon})).
As was already mentioned above we may look for ``radially excited''
global monopoles bifurcating with the de Sitter solution at the other zero
modes $\varphi_K$ for $K>0$. Since the functions $\varphi_K$  have 
$K$ zeros, 
we have to look for solutions of eqs.(\ref{fequs}) with $\baf$'s with 
the same property and and a cosmological horizon. 
The results of a numerical
investigation for the cases $K=1,2$ are summarized in Fig.(\ref{acurves}),
in which the values of the fine-tuning parameter $a$ are plotted as a
function of $\bet^2$. The graphs of the functions $a(\bet^2)$ look very
similar to the one for the case $K=0$ given in \cite{MaisonL} and approach
the latter for $\bet\to 0$.

There is an important difference in the global behaviour between the
fundamental monopole and the radially excited solutions. Whereas the first
ones extend to arbitrarily large values of $r$ with $\baf\to\bet$
and $\mu\to 1-2\bet^2$, the latter
ones have an ``equator'' (maximum of $r$) just outside their cosmological horizon
and then run back to $r=0$, where they become singular. 
The behaviour beyond the equator is quite similar to
the one of ``hairy'' black holes inside their horizon \cite{BLM}. Obviously
$\mu$ cannot tend to $1-2\bet^2$ for $\bet^2<1/2$ without becoming positive
again. This, however, could only be achieved with finite $\psi$ by
further fine-tuning, for which there is no parameter left for given $\bet$.  
One might try to also fine-tune $\bet$ in order find such solutions, but
then one would still need another parameter to suppress the divergent mode of
the Higgs field  for $r\to\infty$ to enforce $\baf\to\bet$.

Since the Schwarzschild coordinates become singular at the equator, 
where $r$ is stationary one has to use a different radial coordinate.
A convenient choice is to take the geodesic distance $s$ in the radial 
direction and determine $r(s)$ solving
\be                                                  \label{geod}
\frac{d}{ds}r=\sqrt{|\mu|}\,.
\ee
Fig.(\ref{slns}) shows the functions $\baf(x)$, $\mu(x)$ and $r(x)$ 
for particular values of $\bet$, where the coordinate $x$ is 
equal to $s$ up to a normalisation factor chosen
such that the cosmological horizon where $\mu$ vanishes 
is situated at $s=1$. The equator at a second zero of $\mu$ is very
close to, but slightly outside the horizon. 
Since $\mu$ stays very small between its two zeros we have inserted
enlargements in Fig.(\ref{slns}) in order to make this behaviour  
visible.

\section{Black holes with global monopole hair}\label{black}

Next we turn to black holes sitting inside global monopoles. Such solutions 
have been described already by S.~Liebling \cite{Liebling}. According to his
results there is a 2-parameter family of such solutions, parametrized by
the radius $\rh$ of their bh-horizon and $\bet$. Encouraged by the results
of the previous section,  we also look for black holes sitting 
inside radially excited global monopoles. 
Indeed, such solutions are readily found numerically solving 
eqs.(\ref{fequs}) with bh boundary conditions eq.(\ref{bhbd}) and 
analogous ones for a cosmological horizon at some $\rc>\rh$.  

In view of the experience with hairy black holes studied in the literature
\cite{BFM}, we expect a limited domain of existence in the $\bet-\rh$-plane 
for such solutions. The numerical analysis shows that there is a qualitative
difference  
between the fundamental solutions (without zeros of the Higgs field) and the 
radially excited ones. Let us first discuss the fundamental solutions.

\begin{figure}
\hbox to\hsize{
  \epsfig{file=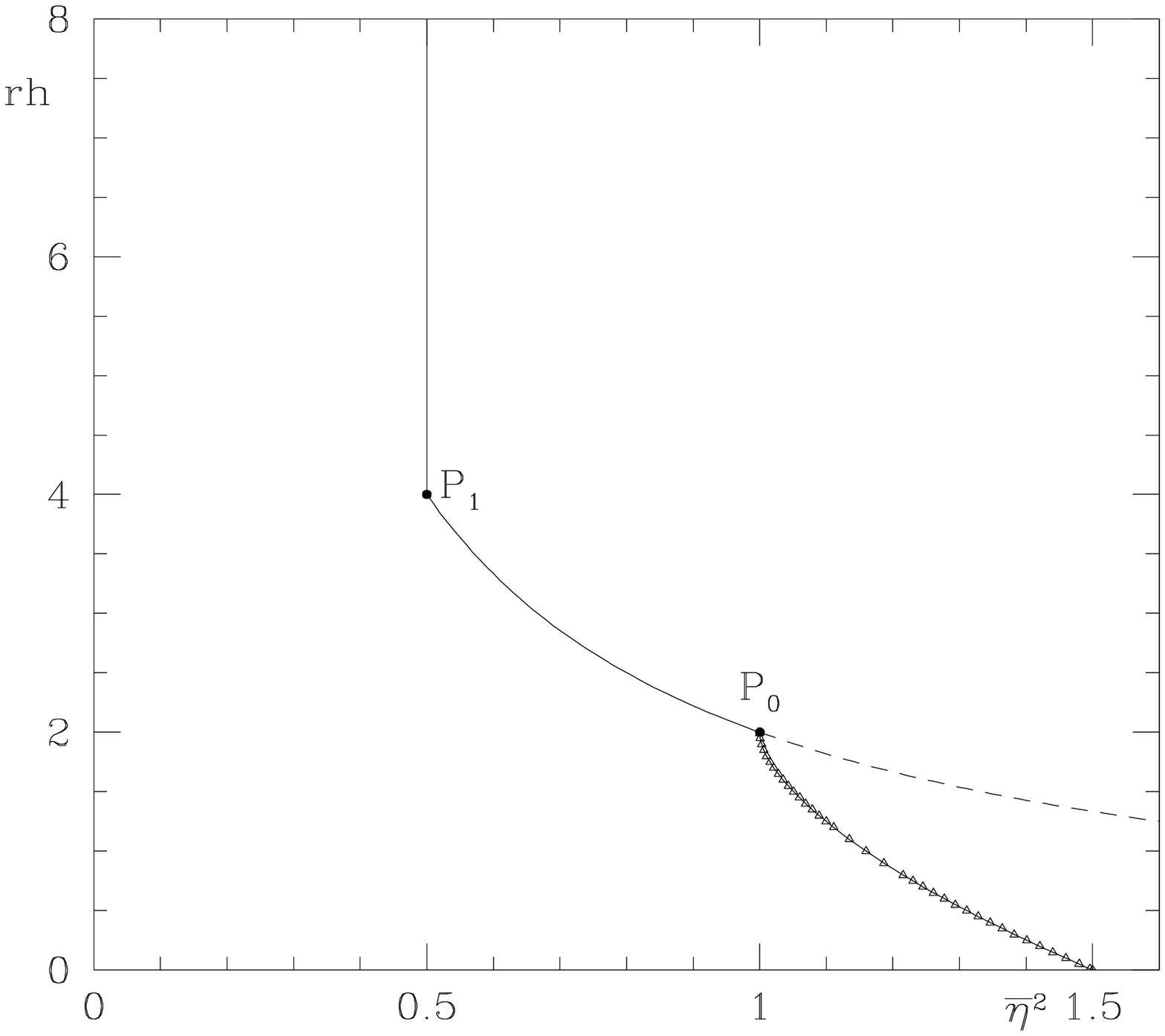,width=0.48\hsize,%
      bbllx=0.7cm,bblly=6.0cm,bburx=20.5cm,bbury=20.0cm}\hss
  \epsfig{file=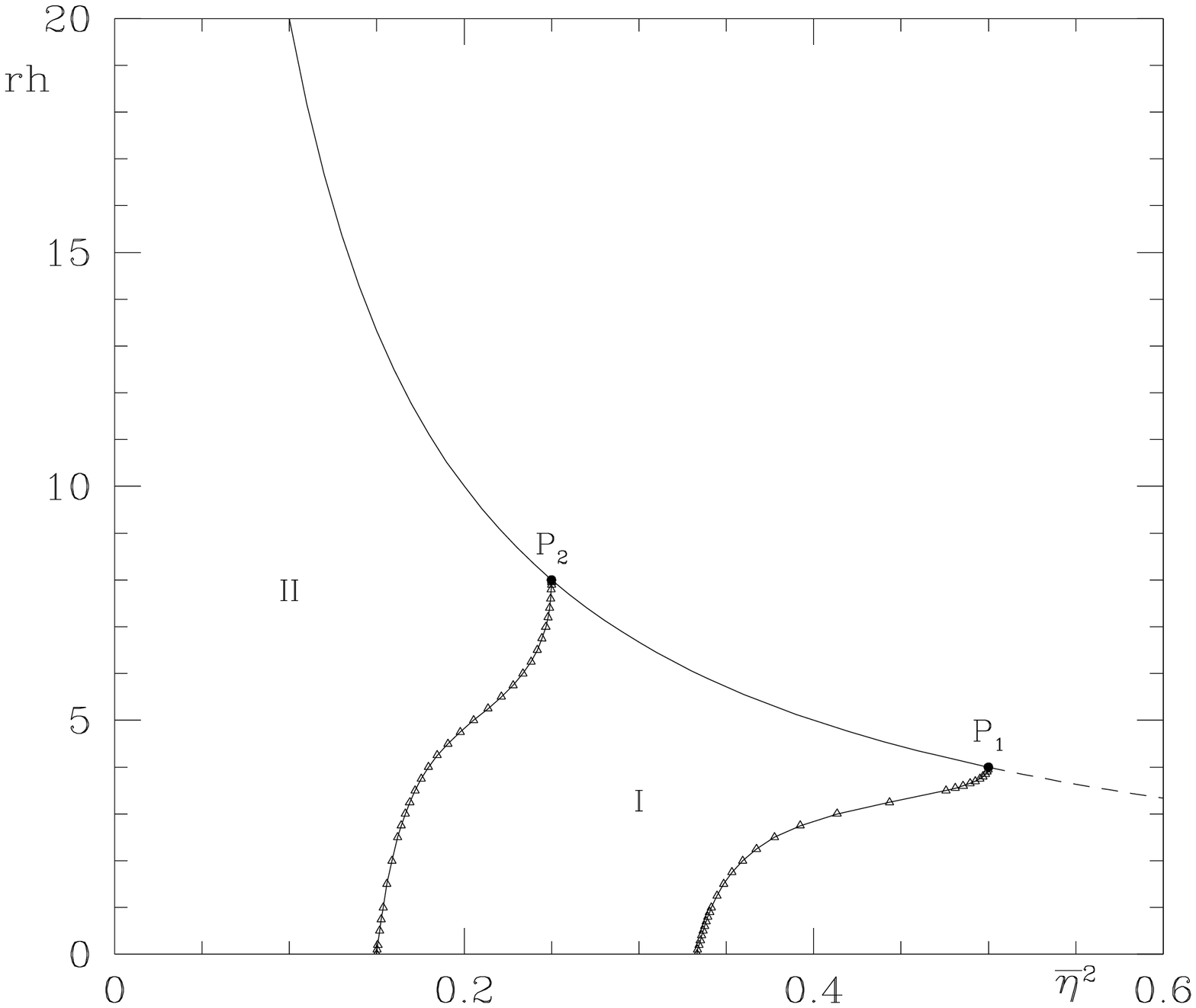,width=0.48\hsize,%
      bbllx=0.7cm,bblly=6.0cm,bburx=20.5cm,bbury=20.0cm}
  }
\caption[domain]{\label{domain}Existence domains for black holes. The first
plot refers to the fundamental solution, the second one to the first and second
radial excitations. The first radial excitation exists in the union of
regions I and II, whereas the second one exists only in region II.  
}
\end{figure}

As shown in Fig.(\ref{domain}) there three regions in the $\bet-\rh$-plane.
For $0<\bet^2<1/2$ black holes with arbitrarily large radius $\rh$ seem to 
exist.
In the interval $1/2<\bet^2<1$ the $\rh$ values are bounded from above 
by the curve $\rh=2/\bet^2$.  
Approaching this curve from below the 
solutions bifurcate with the SdS solution (i.e. $\baf\to 0$), such that at
the same time $\rh$ tends to $\rc$. Comparing with eq.(\ref{chor}) this
yields the relation $\rh=\rc/\sqrt{3}=2/\bet^2$.

In the interval $1<\bet^2<3/2$ the existence domain is bounded by some
curve joining the points $(1,2)$ and $(\sqrt{3/2},0)$ in the
$\bet-\rh$-plane. Approaching this curve the solution again bifurcates
with the SdS solution, but this time with $\rh<\rp$. As in the case of the
regular solution the bifurcation points are determined by the requirement
that a bounded zero mode of the SdS solution exists. The equation to be
solved for $\varphi=r\delta\baf$
is (using rescaled radial variables $x=r/\rc$ and $\xh=\rh/\rc$) 
\be\label{zero}
\frac{d}{dx}(\mu_{\rm SdS}\frac{d}{dx}\varphi)=
     \Bigl(\ov2{x^2}-2-\ov6{\bet^2}-\ov{\xh^3-\xh}{x^3}\Bigr)\varphi\,
\ee
with bh boundary conditions at $x=\xh$ and $\mu_{\rm SdS}$ as given by 
eq.(\ref{SdS}). The requirement for $\varphi$ to stay bounded
for $x\to \rp/rc$ then determines the boundary curve $\rh(\bet)$ shown in
Fig.(\ref{domain}) joining the points $P_0$ and $Q_0$.

The situation is quite similar for the radially excited solutions, the only
difference being the absence of the unbounded piece of the domain.
The existence domains for solutions with one and two zeros of $\baf$
are shown in Fig.(\ref{domain}). The intersection points $P_i, i=0,1,2$, 
etc.\ of the respective boundary curves can be determined analytically as
follows. 

From what was said above they are determined by the condition that
$\rh\to\rp$ as one approaches them along the curve determined by the zero
mode condition. In order to study this limit we assume
$\xh=1/\sqrt{3}-\epsilon$ with $\epsilon<<1$ and introduce a rescaled
variable $y$ defined by
$x=\ov{1}{\sqrt{3}}+\epsilon y$. Keeping only the leading terms as
$\epsilon\to 0$ we obtain from eq.(\ref{zero})
\be
\frac{d}{dy}\Bigl((1-y^2)\frac{d}{dy}\varphi)=
     2(1-\ov{1}{\bet^2})\varphi\,.
\ee
When $x$ varies from $\xh$ to $\rp/rc$ the variable $y$ runs from $-1$ to
$+1$. The bounded solutions on this interval are the Legendre polynomials
$P_K(y)$ obtained for
\be\label{intersec}
\bet_K^2=\ov{2}{K(K+1)+2} \qquad K=0,1,...
\ee     
The first three values are $\bet^2=1,1/2,1/4$ with corresponding values
$\rh=2,4,8$ matching exactly the points $P_0, P_1$ and $P_2$ of
Fig.(\ref{domain}). 

\section{Summary}\label{summ}
In section \ref{global} we put forward some arguments for the existence 
of radial excitations of the static gravitational global monopoles 
possessing a de Sitter like cosmological horizon studied
in \cite{Liebling,MaisonL}. We present some numerical evidence for
the existence of solutions with up to two zeros of the Higgs field.
In section \ref{black} we study corresponding static hairy black hole solutions,
representing black holes sitting inside a global monopole core. In
particular, we determine their existence domains as a function of 
their horizon radius $\rh$.

In a forthcoming publication we shall consider  generalisations of these results
to gravitational monopoles with a dynamical YM field.  

\section{Acknowledgments}
I am indebted to P. Breitenlohner and P. Forg\'acs
for frequent discussions on the subject.

\end{document}